\documentclass[aps,prl,amsmath,two column,amssymb,floatfixng,showpacs,
superscriptaddress,footinbib]{revtex4}
\usepackage[dvips]{graphics}
\usepackage{bm}
\usepackage{float}
\usepackage{epsfig}
\usepackage{enumerate}
\usepackage{amsmath}
\usepackage{color}
\usepackage{graphicx}
\usepackage[colorlinks=true,linktoc=page,linkcolor=red,citecolor=blue]{hyperref}

\newcommand\bea{\begin{eqnarray}}
\newcommand\eea{\end{eqnarray}}
\newcommand\beq{\begin{equation}}  
\newcommand\eeq{\end{equation}}

\begin{document}

\title{\titlename}
\date{\today}
\title{Time-like Entanglement Entropy: a top-down approach}
	\author{Carlos Nunez}
	\email{c.nunez@swansea.ac.uk}
		\affiliation{Department of Physics, Swansea University,
Swansea SA2 8PP, United Kingdom}
\author{Dibakar Roychowdhury}
\email{dibakar.roychowdhury@ph.iitr.ac.in}
\affiliation{Department of Physics, Indian Institute of Technology Roorkee
Roorkee 247667, Uttarakhand,
India}

\date{\today}
\begin{abstract}
We investigate the concept of time-like entanglement entropy (tEE) within the framework of holography.
We introduce a robust top-down prescription for computing tEE {using the holographic duals to} higher-dimensional QFTs—both conformal and confining—eliminating the ambiguities typically associated with analytic continuation from Euclidean to Lorentzian signatures. We present accurate analytic approximations for tEE and time-like separations in slab geometries. We establish a clear stability criterion for bulk embeddings and demonstrate that tEE serves as a powerful tool for computing CFT central charges, extending and strengthening previous results. Finally, we apply our framework to holographic confining backgrounds, revealing distinctive behaviors like phase transitions.
\end{abstract}

\maketitle

\section{Introduction and General Idea}
Maldacena's conjecture \cite{Maldacena:1997re}, along with its refinements \cite{Gubser:1998bc, Witten:1998qj}, introduced the idea that space can emerge from the strongly coupled dynamics of a Quantum Field Theory (QFT). One compelling approach to understanding this emergence of space is through Entanglement Entropy (EE), computed holographically as proposed by Ryu and Takayanagi in \cite{Ryu:2006bv, Ryu:2006ef}; see also \cite{Hubeny:2007xt}. In particular, the works \cite{Swingle:2009bg, VanRaamsdonk:2010pw} clearly suggest how space-like coordinates may emerge from a QFT. This naturally raises the question of whether a time-like coordinate could also emerge from strong dynamics.

With this motivation (among others), the concept of time-like entanglement entropy (tEE) was introduced in \cite{Doi:2022iyj, Doi:2023zaf}. {From the holographic point of view, the notion of tEE was originally introduced in order to understand the emergence of time from quantum entanglement in time-like subsystems in QFTs \cite{Doi:2022iyj, Doi:2023zaf}, much similar in spirit to the original Ryu and Takayanagi prescription \cite{Ryu:2006bv, Ryu:2006ef} where space emerges from quantum entanglement among space-like subsystems. Unlike the usual EE for space-like separated intervals, the analogue of the reduced density matrix (also known as the transition matrix) for time-like separated events is non-Hermitian. Although the computation of tEE in higher dimensional CFTs is a nontrivial task, for 2d CFT this has been obtained following a Wick rotation of the subregion ($A$) associated with the Hilbert space ($\mathcal{H}_A$) \cite{Doi:2023zaf}.} The topic has since become an active area of research, especially within the holographic context. Among the works most influential to our study are \cite{Heller:2024whi, Das:2023yyl, Grieninger:2023knz, Chu:2019uoh, Afrasiar:2024lsi, Afrasiar:2024ldn, Milekhin:2025ycm, Li:2022tsv, Guo:2024lrr, Guo:2025pru, Roychowdhury:2025ukl}.

In the original studies and subsequent developments, the {holographic} computation of time-like entanglement entropy involves extremizing a surface in the Euclidean regime (where it coincides with the standard RT-EE), followed by analytic continuation to Lorentzian time. However, this analytic continuation remains conceptually challenging for higher-dimensional QFT$_d$ ($d \geq 3$), particularly in slab or spherical regions.

One of the goals of the present work is to address this challenge and offer an alternative prescription for computing time-like entanglement entropy {using the holographic prescription for duals to} higher-dimensional QFTs—whether conformal or not. Our approach provides a natural transition from Euclidean to Lorentzian (or time-like) entanglement entropy and establishes a framework for mapping QFT features in a real-time formalism. Another achievement of this work is the derivation of analytic expressions that approximate the time-like EE and time-like separation for slab regions, while also offering a criterion for stability and a computable method for predicting potential phase transitions.

We further present formulas using tEE to compute central charges of CFTs, following the approach of Liu and Mezei \cite{Liu:2012eea, Liu:2013una}. In addition, we explore two holographic duals of confining QFTs and analyse the tEE and its associated phase transitions in these models.

In the remainder of this section, we recall well-established formulas for the strip time-like entanglement entropy, closely following those derived for Wilson loops (see, for instance, \cite{Nunez:2009da}). We also provide analytic expressions that approximate the time-like EE and the time-like separation, and we formulate a criterion for the stability of the entangling surface in the bulk.
\\
{Consider a string background (the extension to eleven dimensions is immediate), holographically dual to a QFT in $d$ space-time dimensions. The string frame metric and dilaton read (there are Ramond and Neveu-Schwarz fields that we do not quote),
\begin{eqnarray}
& &ds_{st}^2= f(u,\vec{y})\Bigg[\lambda dt^2+ d\vec{x}_{d-2}^2+ dv^2+ g(u) du^2 \Bigg]\nonumber\\
& &~~~~~~~~~~~~~ + g_{ij,(9-d)}(u,\vec{y})dy^idy^j,~~\Phi(u,\vec{y}).\label{eq1z}
\end{eqnarray}
We introduced the parameter $\lambda=\pm 1$ to indicate Euclidean metrics, with isometry $SO(d)$ or Lorentzian ones, with isometry $SO(1,d-1)$. This assumption can be relaxed and is discussed in the confining models studied below.
We embed an eight-surface $\Sigma_8$ parameterised by the coordinates $[\vec{x}_{d-2}, \vec{y}_{9-d}, u]$, with $v=\text{constant}$ and $t=t(u)$.
\\
The induced metric of the eight-manifold is,
\begin{eqnarray}
& &ds_{\Sigma_8}^2= f(u,\vec{y})\Bigg[ d\vec{x}_{d-2}^2+ ( g(u)+ \lambda t'^2) du^2 \Bigg]\nonumber\\
& &~~~~~~~~~~+~g_{ij,(9-d)}dy^idy^j.\nonumber\\
& &\det[g_{\Sigma_8}]= f(u,\vec{y})^{d-1}\det[g_{9-d}](g(u)+ \lambda t'^2).\label{eq2z}\
\end{eqnarray}
We calculate the time-like entanglement entropy for a strip (or slab) region as,
\begin{eqnarray}
& & S_{tEE}=\frac{1}{4 G_N}\int d^{9-d}y d^{d-2}x ~du \sqrt{e^{-4\Phi} \det g_{\Sigma_8}}.\label{eq3z}\\
& & S_{tEE}=\frac{\int d^{d-2}x}{4G_N} \int du \times\nonumber\\
& &\int d^{9-d}y\sqrt{e^{-4\Phi(u,\vec{y} ) }  f(u,\vec{y})^{d-1}\det[g_{9-d}](g(u)+ \lambda t'^2). } \nonumber
\end{eqnarray}
The evaluation of the integral over the $(d-2)$ coordinates represented by $\vec{x}$  is immediate (as they are isometric coordinates). Less trivial could be the evaluation over the $(9-d)$ coordinates represented by $\vec{y}$. It usually occurs that the expression $e^{-4\Phi(u,\vec{y} ) }  f(u,\vec{y})^{d-1}\det[g_{9-d}]$ factorises into a function of $\vec{y}$ times a function of the radial coordinate $u$. This is the case in the examples discussed below and in those studied in the forthcoming work \cite{NRtoappear}. After evaluating the integrals over the seven coordinates $(\vec{y},\vec{x})$ we arrive at a generic expression for the time-like entropy on slab-regions,
}
%
\begin{eqnarray}
 & & S_{tEE}=\frac{{\cal N}}{4 G_{10}}\int_{u_0}^\infty du \sqrt{G^2 + F^2 t'^2(u)},\label{liostrip}
 \end{eqnarray}
 Here $\cal N$ is a constant, $F,G$ are functions of the radial coordinate $u$.
 
 One can define a function ${\cal V}$ in terms of which the separation between the two regions is expressed. The function and the time-like separation in the case of a strip are
 \begin{eqnarray}
 & & {\cal V}=\frac{F(u)}{G(u) F(u_0)}\sqrt{F^2(u)-F^2(u_0)} ,\label{liostrip2}\\
 & & T=2\int_{u_0}^\infty \frac{du}{\cal V}.\nonumber\end{eqnarray}
 Using a first integral of the equations of motion derived from eq.(\ref{liostrip}),  the time-like entanglement entropy reads,
 \begin{eqnarray}
 & & \frac{4 G_{10}}{\cal N}S_{tEE}=
 F(u_0) T + \nonumber\\
 & &
 2 \int_{u_0}^\infty du \frac{G(u)}{F(u)}\sqrt{F^2(u)- F^2(u_0)} - 
2\int_{u_*}^\infty G(u) du \nonumber\\
& & =2\int_{u_0}^\infty\!\!  du \frac{F(u) G(u)}{\sqrt{F^2(u)- F^2(u_0)}}- 2\int_{u_*}^\infty\!\! G(u) du.\label{liostrip3}
\end{eqnarray}
Usually, analytically evaluating the integrals in eqs.(\ref{liostrip2})-(\ref{liostrip3}) is not possible. There are approximate expressions that are very useful. For the time separation $T$ in the case of the strip the expression proposed in \cite{Kol:2014nqa} reads,
\begin{equation}
 T_{app}=\frac{\pi G(u)}{F'(u)}\Bigg|_{u=u_0}. \label{Tappeq}
\end{equation}
A useful quantity indicating whether the embedding is stable under fluctuations was proposed in \cite{Faedo:2013ota}. This quantity is
\begin{equation}
 Z(u_0)=\frac{d }{du} \left(\frac{\pi G(u)}{F'(u)} \right)\Big|_{u=u_0}.\label{Zdeu}  
\end{equation}
In fact, stability of the embedding is implied by $Z(u_0)<0$, on the other hand for positive $Z(u_0)$ the embedding is unstable \cite{Faedo:2013ota}. {This criterium for stability can be proven using the formal similarity between the action for a generic Wilson loop and that of the time-like EE in eq.(\ref{liostrip}). The proof is presented in \cite{NRtoappear}. Intuitively, it can be understood as follows: if the time-separation is not monotonically increasing towards smaller values of $u_0$, there is a value of $u_0$ for which there are two possible time-separations. Characteristically, this leads to a competition between embeddings and a phase transition. This is encoded by the derivative of the time separation turning positive $Z(u_0)>0$. We discuss examples of this phase transition in the confining models discussed below. }

For the approximate time-like entanglement entropy of a strip we present an expression that relies on eq.(\ref{Tappeq}). The derivation is presented in \cite{NRtoappear},
\begin{eqnarray}
& & \frac{4G_{10}}{\cal N} S_{tEE,app}(u_0)  =\int^{u_0} dz ~F(z)T_{app}'(z)\label{StEEappeq}\\
& & =\!\! \int^{u_0}\!\! dz \frac{F(z)}{F'(z)^2} \left[ G'(z) F'(z)- G(z) F''(z) )\right].\nonumber
\end{eqnarray}
 Using these expressions, let us  study the time-like entanglement entropy for generic CFTs in $d$-dimensions.

\section{Time-like Entanglement Entropy for generic CFTs}
We develop a top-down approach to derive expressions for the time-like entanglement entropy and for the time separation. We reproduce and generalise expressions already derived using a bottom-up approach \cite{Doi:2022iyj, Doi:2023zaf} and give a nice interpretation to the result in \cite{Doi:2023zaf}. We present expressions for the time-like entanglement entropy and the time-separation, both  for the cases of slabs and for spherical entangling regions.
\\
Let us consider a family of backgrounds dual to  generic CFTs in $d$-dimensions.
\begin{align}
    ds^2_{10}=f_1(y)ds^2_{AdS_{d+1}}+g_{ij}(y)dy^i dy^j,\label{metricgeneral}
\end{align}
where $g_{ij}$ is the metric of the internal manifold and
\begin{equation}
    ds^2_{AdS_{d+1}}|_{\Sigma^{(\lambda)}}=u^2(\lambda dt^2+dv^2+d \textbf{x}_{d-2})+\frac{du^2}{u^2},\label{betoalonso1}
\end{equation}
that is used to compute the tEE in strip regions. For spherical/hyperbolic regions we write the AdS metric as,
\begin{equation}
ds^2_{AdS_{d+1}}|_{\hat{\Sigma}^{(\lambda)}}=u^2(\lambda dt^2+t^2 d\Omega^{(\lambda)}_{d-2}+dv^2)+\frac{du^2}{u^2}.\label{betoalonso2}
\end{equation}
We use the parameter $\lambda=\pm 1$ with $(+)$ to indicate the Euclidean case and $(-)$ the Minkowski signature. This parameter keeps track of the character of the calculation. For $\lambda=+1$ $\Omega^{(\lambda)}$ denotes a sphere whilst for $\lambda=-1$, it denotes the compact part of a hyperbolic space.


In each of the above cases the eight manifold needed to calculate the time-like entanglement entropy is parametrised by setting $v=0$ in eqs.(\ref{metricgeneral})-(\ref{betoalonso2}), and considering an embedding of the form $t=t(u)$. The  line elements and expressions for the separation $T$ and the tEE are studied below in each case.

\underline{{\bf The case of the strip}}. 
The  line element of the eight-manifold is given by
\begin{eqnarray}
 & &   ds^2_8|_{\Sigma_8^{(\lambda)}}=f_1(y)(1+\lambda u^4 t'^2(u))\frac{du^2}{u^2}+f_1(y)u^2 d \textbf{x}_{d-2}+\nonumber\\
    & & g_{ij}(y)dy^i dy^j 
\end{eqnarray}
The time-like entanglement entropy reads
\begin{eqnarray}
\label{e74}
 & &   S^{(\lambda)}_{EE}[\Sigma_8^{(\lambda)}]=\frac{1}{4G_{10}}\int d^8 x \sqrt{e^{-4 \Phi}\det g_8}= \nonumber\\
 & & \frac{\mathcal{N}}{4G_{10}}\int_{u_0}^{\infty}du u^{d-3}\sqrt{1+\lambda u^4 t'^2(u)}.
\end{eqnarray}
The time-like entanglement entropy arises when $\lambda=-1$. For $\lambda=1$, we have the usual Ryu-Takayanagi entanglement entropy \cite{Ryu:2006bv}. In eq.(\ref{e74}) we denoted
\begin{eqnarray}
    & & \mathcal{N}=L^{(d-2)}\int d^{9-d}y \sqrt{e^{-4\Phi}\det g_{ij}}f_1^{\frac{d-1}{2}}(y)~\nonumber\\
    & & L^{(d-2)}=\int d^{d-2}x.\label{paredes}
\end{eqnarray}

The expression in eq.\eqref{e74} has a generic structure of the form in eq.(\ref{liostrip}), with
\begin{equation}
  G=u^{d-3} ~~~\text{and}~~~ F=\sqrt{\lambda}u^{d-1} . 
\end{equation}
 Following the expressions in eq.(\ref{liostrip2}), we find
\begin{align}
    {\cal V}=
    \frac{\sqrt{\lambda}u^2}{u^{d-1}_0}\sqrt{u^{2d-2}-u^{2d-2}_0}.
\end{align}
The time separation is
\begin{eqnarray}
\label{e79}
  & &   T = \frac{2}{\sqrt{\lambda}}u_0^{d-1}\int_{u_0}^\infty \frac{du}{u^2}\frac{1}{\sqrt{u^{2d-2}-u^{2d-2}_0}}\nonumber\\
 & & = 
 \frac{2 \sqrt{\pi}~ \Gamma\left( \frac{d}{2d-2}\right)}{\sqrt{\lambda}~ \Gamma\left(\frac{1}{2d-2} \right)} \times\frac{1}{u_0}   .
\end{eqnarray}
Note that the time separation is purely imaginary for Minkowski signature ($\lambda=-1$). In contrast to \cite{Doi:2023zaf}, we {\it do not} continue $T\to iT$, all the effect of calculating with Lorentzian signature is encoded by ${\lambda}$. The approximate expression  in eq.(\ref{Tappeq}) for $T_{app}$ captures the behaviour of the function $T(u_0)$. In fact, using eq.(\ref{liostrip}),
\begin{align}
\label{e80}
    T_{app}=\pi \frac{G}{F'}\Big|_{u=u_0}=\frac{\pi}{\sqrt{\lambda}(d-1)u_0}.
\end{align}
%
%
%
Notice that $Z(u_0)$ defined in eq.(\ref{Zdeu}) is always negative indicating that all these embeddings are  stable.

We  compute the tEE and its approximate expression, using eqs.(\ref{liostrip}) and (\ref{StEEappeq}). We find,
\begin{eqnarray}
  & &   \frac{ 4G_{10} }{\cal N }S^{(\lambda)}_{EE}[\Sigma_8^{(\lambda)}] =\label{Steeu0}\\
  & & \,\left[\frac{2}{(2-d)} {}_2F_1\left(\frac{1}{2},\frac{2-d}{2 d-2};\frac{d}{2 (d-1)};1\right)\right]\times u_0^{d-2}=\nonumber\\
  & & \frac{1}{(2-d)} \left[ 2\sqrt{\pi} \frac{\Gamma\left( \frac{d}{2d-2}\right)}{\Gamma\left(\frac{1}{2d-2} \right)}\right]^{(d-1)} \frac{1}{\sqrt{\lambda^{d-2}} ~~|T|^{d-2}}.\nonumber
\end{eqnarray}
We have used eq.(\ref{e79}) to replace $u_0$ in terms of $T$. The expression of eq.(\ref{Steeu0}) is an exact and regularised expression, valid for $3\leq d$.
We compare this expression with those in equations (4.36)-(4.37) in the paper \cite{Doi:2023zaf}, obtaining agreement.  {Note that depending on the space-time dimension $d$, one may have imaginary values for the tEE when written in terms of the time separation $T$. We believe this imaginary value should be considered physical. In special circumstances this imaginary value can be understood in field theoretic terms, see \cite{Guo:2024lrr, Guo:2025pru}.} 

In \cite{Doi:2023zaf}
the Newton constant in $(d+1)$-dimensions is used, here this that appears in the quotient $\frac{1}{G_{d+1}}=\frac{\cal N}{G_{10}}$. As we explain below, the effective Newton constant in lower dimensions depends on the particular dual CFT and it is related to the central charge of such dual CFT.

For Lorentzian signature ($\lambda=-1$), the quantities $T,T_{app}$ are purely imaginary. The time-like entanglement entropy in terms of $T$ is real for even-d  and purely imaginary for odd-d (we remind the reader that $d$ is the dimension of the holographically dual QFT). We  calculate the approximate-tEE using eq.(\ref{StEEappeq}). We find,
\begin{equation}
S_{app}= \!=\! \int^{u_0} dz ~F(z)T_{app}'(z)= -\frac{\pi}{(d^2-3d+2)} u_0^{d-2}. \label{e16}   
\end{equation}
This expression should be supplemented by an integration constant, that we omitted above. The same logic for the Lorentzian result applies.

The result in eq.(\ref{e16}) agrees with the exact behaviour in terms of the turning point $u_0$. Combining  eqs.(\ref{e80}) and (\ref{e16}) we find,
 \begin{align}
       S_{app}=-\frac{\pi^{d-1}}{\sqrt{\lambda^{(d-2)}}(d-1)^{d-2}(d^2-3d+2)}\frac{1}{|T_{app}|^{d-2}}.
   \end{align}
Again, we note that for Minkowski signature ($\lambda=-1$) the approximate time-like entanglement in terms of the approximate time separation is { real for even dimensions and purely imaginary in odd dimensions.} 

Let us now replace the slab/strip by a sphere or a hyperboloid.

\underline{{\bf The case of the sphere/hyperboloid}}. 
After setting the coordinate $v=0$ and $t=t(u)$ in eqs.(\ref{metricgeneral}), (\ref{betoalonso2}), the corresponding line element of the eight manifold needed to calculate tEE  is
\begin{eqnarray}
 & &     ds^2_8|_{\hat{\Sigma}_8^{(\lambda)}}=f_1(y)(1+\lambda u^4 t'^2(u))\frac{du^2}{u^2}+f_1(y)u^2 t^2 d \Omega^{(\lambda)}_{d-2}\nonumber\\
 & & +g_{ij}(y)dy^i dy^j. 
\end{eqnarray}
The tEE is given by
\begin{eqnarray}
 & &   S^{(\lambda)}_{EE}[\hat{\Sigma}_8^{(\lambda)}]=\frac{1}{4G_{10}}\int d^8 x \sqrt{e^{-4 \Phi}\det g_8}\nonumber\\
 & & =\frac{\widehat{\mathcal{N}}}{4G_{10}}\int_{u_0}^\infty du u^{d-3}t^{d-2}\sqrt{1+\lambda u^4 t'^2(u)},\label{cani}
\end{eqnarray}
where we denote
\begin{align}
    \widehat{\mathcal{N}}=\text{Vol}(\Omega^{(\lambda)}_{d-2})\int d^{9-d}y \sqrt{e^{-4\Phi}\det g_{ij}}f_1^{\frac{d-1}{2}}(y).\label{cutiromero}
\end{align}
The equation of motion that follows from the 'action' in eq.(\ref{cani}) is solved by
\begin{align}
\label{e89}
    t(u)= \frac{\sqrt{R^2 u^2-\lambda }}{u}\;.
\end{align}
We introduce a small parameter $\epsilon$ to regulate UV-divergencies and change variables according to $u=\frac{\sqrt{\lambda}}{R} x$. Using eq. (\ref{e89}), the tEE in eq.(\ref{cani}) reads,
\begin{align}
 \frac{4G_{10} S^{(\lambda)}_{EE}[\hat{\Sigma}_8^{(\lambda)}]}{
 \widehat{\mathcal{N}}\lambda^{(d-2)/2 }} =\int_1^{\frac{R}{\sqrt{\lambda}\epsilon}}dx (x^2-1)^{\frac{d-3}{2}}.
\end{align}
Let us evaluate this integral. For odd dimension $d$ the result is--see \cite{Jokela:2025cyz}
\begin{eqnarray}
 & &    \frac{ 4G_{10} S^{(\lambda)}_{EE}[\hat{\Sigma}_8^{(\lambda)}]}{{\widehat{\mathcal{N}}\lambda^{(d-2)/2}}{}}=\sum_{j=0}^{[\frac{d-3}{2}]}\frac{\Big( \frac{3-d}{2}\Big)_j}{j! (d-2j -2)}\Big(\frac{R}{\sqrt{\lambda}\epsilon} \Big)^{d-2j-2}\nonumber\\
    & & -(-1)^{\frac{d+1}{2}}\frac{\sqrt{\pi}
    \Gamma\left(\frac{(d-1)}{2}\right)}{2\Gamma \left(\frac{d}{2} \right)}.\label{StEEodd}
\end{eqnarray}
We used the Pochhammer symbol $\Big( \frac{3-d}{2}\Big)_j=\frac{\Gamma \Big(\frac{3-d}{2}+j\Big)}{\Gamma \Big(\frac{3-d}{2}\Big)}$.
\\
For even $d$, we follow \cite{Jokela:2025cyz} to obtain,
\begin{eqnarray}
  & &   \frac{ 4G_{10}~S^{(\lambda)}_{EE}[\hat{\Sigma}_8^{(\lambda)}]}{{\widehat{\mathcal{N}}\lambda^{(d-2)/2}}}=\sum_{j=0}^{[\frac{d-3}{2}]}\frac{\Big( \frac{3-d}{2}\Big)_j}{j! (d-2j -2)}\Big(\frac{R}{\sqrt{\lambda}\epsilon} \Big)^{d-2j-2}\nonumber\\
  & & -\frac{\Gamma\left(\frac{d-1}{2} \right)}{\Gamma \left(\frac{d}{2} \right)}\frac{(-1)^{d/2}}{\sqrt{\pi}}\Big(\log (\frac{2R}{\epsilon \sqrt{\lambda}}) +\frac{1}{2}\mathcal{H}_{\frac{d-2}{2}}\Big).\label{StEEeven}
\end{eqnarray}
We denoted by $\mathcal{H}_{n}=1+\frac{1}{2}+\frac{1}{3}+\cdots+\frac{1}{n}$ the  Harmonic numbers.

{For $\lambda=+1$, this result coincides with that written in equations (4.5) and (4.6) of the paper \cite{Doi:2023zaf}. For $\lambda=-1$, this reproduces the bottom-up result of equations (4.9)-(4.10) of \cite{Doi:2023zaf}.} 

Notice that we by-pass the important problem of finding the time-like or space like surface homologous to the time-like subregion. This problem was carefully discussed in \cite{Li:2022tsv}.

In summary, our approach  uses an eight manifold if the holographic background is  in Type II and a nine manifold if it were written in eleven dimensional supergravity. It clarifies the relation between the ten dimensional Newton constant $G_{10}$ and the one on lower dimensional supergravity of the bottom up approach $G_{d+1}=\frac{G_{10} \text{Vol}\Omega_{d-2}}{\widehat{\cal N}}$. We now display a relation between the central charge of the dual CFT
 and $\widehat{\cal N}$. 
 \\
\underline{\bf A Liu-Mezei central charge}
\\
It is natural, given the results in eqs.(\ref{StEEodd}),(\ref{StEEeven}), to apply the Liu-Mezei \cite{Liu:2012eea} formalism to define a central charge. In particular, we define for dimension $d$-odd
\begin{equation}
(d-2)!! c_{LM,odd}= \left(R\partial_R -1 \right) ....\left(R\partial_R - d+2\right) S^{(\lambda)}_{EE}[\hat{\Sigma}_8^{(\lambda)}].\label{LModd}
\end{equation}
For dimension $d$-even, we have
\begin{equation}
(d-2)!! c_{LM,even}= R\partial_R  ....\left(R\partial_R - d+2\right) S^{(\lambda)}_{EE}[\hat{\Sigma}_8^{(\lambda)}].\label{LMeven}
\end{equation}
In both cases, we should take {\it the absolute value of the result} to guarantee a positive number.
We can also define a central charge using the slab time-like entanglement entropy by calculating 
\begin{equation}
 c_{slab}\propto \frac{T^{d-2}}{ L^{d-2}} T \partial_T S_{EE}^{(\lambda)} .\label{eldibu}  
\end{equation}
For related treatment of central charge using the tEE, see the paper \cite{Giataganas:2025div}. As an illustration of the previous results, we test all the expressions derived in this section for an infinite family of four dimensional ${ N}=2$ SCFTs. These checks can be generalised, see \cite{NRtoappear}.
\subsection{An explicit example}
%
%
%
%
%
%
%
To illustrate our expressions, let us discuss the tEE for ${ N}=2$ SCFTs in four dimensions. The SCFTs of choice for this example are long linear quivers (of length $P\to\infty$). See the papers \cite{Gaiotto:2009gz,Aharony:2012tz,Reid-Edwards:2010vpm, Lozano:2016kum, Nunez:2019gbg, Nunez:2018qcj, Macpherson:2024frt} for a summary of field theory aspects and to set the notation. The background consists of a metric, a dilaton, NS $H_3$ field and Ramond $F_2,F_4$ fields. For our calculation, only the metric and dilaton are needed. These  read (in a convention where $g_s=\alpha'=1$),
\begin{eqnarray}
& & ds^2_{10}=\sqrt{\tilde{f}^3_1 \tilde{f}_5} \Big[4ds^2_{AdS_5}+\tilde{f}_2 d\Omega_2 (\theta , \phi)+ \tilde{f}_3 d \chi^2 +
\nonumber\\
&  &
~~~~~~~~~~~~~\tilde{f}_4 (d \sigma^2 + d \eta^2)\Big], ~~~e^{-4 \Phi}=(\tilde{f}_1 \tilde{f}_5)^{-3}.
\end{eqnarray}
The functions $\tilde{f}_i(\sigma,\eta)$ are expressed in terms of a single function $V(\sigma,\eta) $ as \cite{Nunez:2018qcj}-\cite{Macpherson:2024frt}
\begin{eqnarray}
& & \tilde{f}^3_1 = \frac{\dot{V}\Delta}{2V''},~~\tilde{f}_2=\frac{2 V'' \dot{V}}{\Delta},~~\tilde{f}_3=\frac{4 \sigma^2 V''}{2\dot{V}-\ddot{V}},~~\tilde{f}_4=\frac{2 V''}{\dot{V}},\nonumber\\
& & \tilde{f}_5=\frac{2(2 \dot{V}-\ddot{V})}{\dot{V} \Delta},~~~\Delta = (2 \dot{V}-\ddot{V})V''+(\dot{V}')^2,\nonumber\\
& & \dot{V}=\sigma \partial_\sigma V,~~~\ddot{V}=\sigma\partial_\sigma\dot{V},~~ V'' =\partial^2_\eta V.
\end{eqnarray}
In turn, the function $V(\sigma,\eta)$ can be written in terms 
of the CFT data, encoded in a rank function $R(\eta)$, with Fourier decomposition coefficients $R_k$,
\begin{eqnarray}
& & V(\sigma,\eta)=-\sum_{k=1}^\infty R_k \sin \left( \frac{k \pi\eta}{P}\right) K_0\left( \frac{k \pi \sigma}{P}\right)\nonumber\\
& &~\text{where}~~~R_k=\frac{2}{P}\int_0^P {\cal R}(\eta) \sin \left( \frac{k \pi\eta}{P}\right) d\eta.\label{Vsigmaeta}   
\end{eqnarray}
In order to compute tEE,
we identify the quantities in eqs.(\ref{metricgeneral})-(\ref{betoalonso2}),
\begin{eqnarray}
 & & d=4,~~f_1(y)=4 \left( \tilde{f}_1^3\tilde{f}_3\right)^{1/2},\\
 & & g_{ij}dy^idy^j\!=\!\! \sqrt{\tilde{f}^3_1 \tilde{f}_5}\! \Big[\tilde{f}_2 d\Omega_2 (\theta , \phi)\!+\! \tilde{f}_3 d \chi^2 +
\tilde{f}_4 (d \sigma^2 + d \eta^2)\Big].\nonumber
\end{eqnarray}
Using eqs.(\ref{paredes}),(\ref{cutiromero}) we find
\begin{eqnarray}
& & {\cal N}= 256 \pi^2 L^2 \int_0^\infty d\sigma\int_0^P d\eta ~~\dot{V}V''\sigma,\label{dibu1}\\
& & \widehat{{\cal N}}= 256 \pi^2 \text{Vol}S^2_{\lambda} \int_0^\infty d\sigma\int_0^P d\eta ~~\dot{V}V''\sigma.\label{dibu2}
\end{eqnarray}
Using eq.(\ref{Vsigmaeta}) it follows that
\begin{equation}
\int_0^\infty d\sigma\int_0^P d\eta ~~\dot{V}V''\sigma=\frac{P}{4}\sum_{k=1}^\infty R_k^2,\label{dibu3}    
\end{equation}
from this we find
\begin{equation}
{\cal N}= 64 \pi^2 L^2 P \sum_{k=1}^\infty R_k^2,~~ \widehat{\cal N}= 64 \pi^2 \text{Vol}S^2_{\lambda} P \sum_{k=1}^\infty R_k^2\; . \label{dibu4}
\end{equation}The coefficients
in eqs.(\ref{dibu1})-(\ref{dibu4}) appear when computing the free energy of any member of the family of 4d SCFTs, see for example \cite{ Nunez:2019gbg,Macpherson:2024frt}. In other words the tEE captures (by intermediate of the coefficients ${\cal N}$ and $ \widehat{\cal N}$), the central charge of the dual CFT. This result is in agreement with that obtained via localisation-matrix model approach, see \cite{Nunez:2023loo}.

The expressions for the time-like entanglement entropy and the time separation $T$ in the case of the slab  are,
\begin{eqnarray}
 & &T= \frac{\zeta}{\sqrt{\lambda} ~~u_0},~~\frac{ 4G_{10} }{\cal N }S^{(\lambda)}_{EE}[\Sigma_8^{(\lambda)}] =- \frac{\zeta}{2}u_0^2,~
 \zeta= 2\sqrt{\pi}\frac{\Gamma\left(\frac{2}{3} \right)}{\Gamma\left(\frac{1}{6} \right)}.\nonumber\\
 & &  S^{(\lambda)}_{EE}[\Sigma_8^{(\lambda)}]=-\frac{\cal N}{8 G_{10}}\frac{\zeta^3}{\lambda ~T^2}.\label{taglia}
\end{eqnarray}
Using eq.(\ref{eldibu}), we  find
\begin{eqnarray}
  c_{slab}\propto \frac{{\cal N}~\zeta^3}{4  G_{10}L^2~\lambda}  
\end{eqnarray}
Note that $\frac{\pi}{2} \zeta^3\approx 1$ and $\frac{\cal N}{L^2}$ is proportional to the volume of the internal manifold that appears in eq.(\ref{paredes}). Hence $c_{slab}$ is proportional to the free energy of the dual CFT obtained by purely field theoretical means.
\\
For the tEE on the sphere/hyperbolic space we find,
\begin{eqnarray}
  & &   { S^{(\lambda)}_{EE}[\hat{\Sigma}_8^{(\lambda)}]}\!=\!\frac{\widehat{\mathcal{N}}~\lambda{}}{4G_{10}~}\left(\!
\frac{R^2}{2\lambda \epsilon^2} -\frac{1}{2}\log\left(  \frac{2R}{\epsilon \sqrt{\lambda}}\right) -\frac{1}{4}\!\right).
\end{eqnarray}
Using eq.(\ref{LMeven}), the Liu-Mezei central charge is
\begin{equation}
    c_{LM}=\frac{\lambda}{8 G_{10}}\widehat{\cal N}.
\end{equation}
The Euclidean case ($\lambda=+1$) gives the usual central charge. The Lorentzian case suggests that we should take the {\it absolute value} of the result (as we anticipated), in order to interpret this as a number of degrees of freedom. Note also that $c_{slab}$ and $c_{LM}$ are measuring the same physical observable, related to the free energy of the dual CFT.

The example above can be extended to other dimensions.
In fact, for any AdS$_{d+1}$ (with $3\leq d$) it is feasible to study systems testing the calculations we presented in this section.
The details are given in \cite{NRtoappear}. {It is interesting to observe that the definition for time-like entanglement entropy used in this work is invariant under generic U-duality. For example, we could have considered the eleven dimensional supergravity description of the family of backgrounds in this section. After calculating the induced metric on a nine manifold, the result obtained for the tEE would be the same. In other words, the dimensionality of the space-time $d=4$ does not change, as it refers to the quantum field theory and does not rely on the string or M-theory embedding. This is a virtue of our definition being U-duality invariant.}

Below we leave the conformal realm and discuss the calculation of the tEE for a strip region in the case of a confining QFT.

\section{Time-like Entanglement Entropy for Confining Systems}
There are various established holographic duals to confining field theories, see for example \cite{Witten:1998zw, Klebanov:2000hb, Maldacena:2000yy}. To begin with,
we choose to work with a model proposed by Witten \cite{Witten:1998zw}. The model consist of a stack of D4 branes that wrap a circle. SUSY breaking boundary conditions on the circle are imposed for the fields.
The second type of model is a $(2+1)$ dimensional and SUSY version of the Witten model discussed above. It consists of D3 branes that wrap a circle with periodic boundary conditions for bosons and antiperiodic for fermions. A twist is performed on this field theory that allows the preservation of four supercharges. The system flows from $N=4$ SYM in $(3+1)$ dimensions to a SUSY gapped QFT in $(2+1)$ dimensions. The perturbative spectrum of such theory is discussed in \cite{Kumar:2024pcz, Castellani:2024ial}. The holographic dual was presented in \cite{Anabalon:2021tua} and further studied in \cite{Chatzis:2024top, Chatzis:2024kdu, Chatzis:2025dnu}.
Let us start our study with the system of D4 branes wrapping $S^1$.
\subsubsection{\underline{Witten's model for $(3+1)$ Yang-Mills}}
The  metric and dilaton (there is also a Ramond $F_4$ that we do not quote) are,
\begin{eqnarray}
	& & ds^2_{10}=f_1(u) (\lambda dt^2 + dx^2+dy^2+dv^2)+f_2(u) d\phi^2+ \nonumber\\
    & &\frac{du^2}{f_2(u)}+ f_3(u) d\Omega_4,\nonumber\\
	& & f_1(u) = \frac{u^{3/2}}{R^{3/2}}~;~f_2(u) =f_1(u) h(u)~;~h(u)=1-\frac{u_\Lambda^3}{u^3} ~;\nonumber\\
    & & f_3(u) = \frac{u^2}{f_1(u)},~~
     e^{4\Phi(u)}=g_s^4 f_1^2(u).\label{wittenmodel}
\end{eqnarray}
The eight manifold used to calculate the tEE on a strip is $\Sigma_8=[x,y,\Omega_4,\phi,u]$, with $v=0$ and $t(u)$. We write
\begin{eqnarray}
& & 4 G_{10}S_{tEE,strip}=\int d^8x \sqrt{e^{-4\Phi}\det [g_8]}=\nonumber\\
& &\tilde{\cal N}\int_{u_0}^\infty \sqrt{F^2t'^2+ G^2},~~ G=u,~~F= \sqrt{\lambda}~ u~ f_1(u)~\sqrt{h(u)},\nonumber\\
& & \tilde{\cal N}= \frac{R^3}{g_s^2}\text{Vol}[S^4]L_\phi~ L_x~L_y.\label{teed4}
\end{eqnarray}
Using  eqs.(\ref{liostrip})-(\ref{liostrip3}) we write the time separation $T$ and time-like entanglement entropy,

\begin{eqnarray}
    & & T =  \frac{2R^{3/2}}{\sqrt{\lambda}}u_0 \sqrt{u^3_0 -u^3_\Lambda} \times\nonumber\\
    & &\int_{u_0}^\infty  \frac{du}{\sqrt{u^3-u^3_\Lambda}}\frac{1}{\sqrt{u^2(u^3 -u^3_\Lambda)-u_0^2(u_0^3 -u^3_\Lambda)}}.\label{TEED4}\\
    & &  \frac{ 2 G_{10} S_{EE}}{\tilde{ \mathcal{N}}}=
    2\int_{u_0}^\infty du \frac{u^2 \sqrt{u^3- u^3_\Lambda}}{\sqrt{u^2(u^3 -u^3_\Lambda)-u_0^2(u_0^3 -u^3_\Lambda)}}\nonumber\\
    & &-2\int_{u_\Lambda}^\infty du u.\label{teeD4b}
\end{eqnarray}
Like in the conformal case, the Lorentzian signature $(\lambda=-1)$ case gives a purely imaginary time-separation. We evaluate the approximate expressions for the time separation and tEE in eqs.(\ref{Tappeq})-(\ref{StEEappeq}). We find,
\begin{eqnarray}
    & & T_{app}=\frac{2 \pi  R^{3/2} u_0^{5/2} \sqrt{1 -\frac{  u^3_\Lambda }{u_0^3}}}{\sqrt{\lambda}  \left(5 u_0^3-2u^3_\Lambda \right)}.\label{TEEappD4}\\
    & & Z(u_0)= \sqrt{\frac{\pi^2 R^3}{\lambda (u_0^3- u_\Lambda^3)}}\frac{(-5 u_0^6 + 10 u_0^3 u_\Lambda^3 + 4 u_\Lambda^6)}{(5u_0^3-2 u_\Lambda^3)^2}\label{Zwitten}\\
    & & S_{app}(u_0)= -\pi\frac{u_0^2\left( u_0^3 +2 u_{\Lambda}^3 \right)}{10u_0^3 -4u_{\Lambda}^3}. %
\label{approxtEED4S1}
\end{eqnarray}
As we mentioned above, there should be an integration constant in $S_{app}$, which can be needed to compare with the exact expression (that in this case must be obtained numerically). Notice that the double-valuedness of $T_{app}$ indicates the possibility of a phase transition. The function $Z(u_0)<0$ indicates that the embedding is stable for $u_0^3>u_\Lambda^3(1+ \frac{3}{\sqrt{5}})$, but unstable for values of $u_0$ closer to $u_\Lambda$. These two  indications suggest that the phase transition must take place.

\begin{figure}
    \centering
    \includegraphics[width=0.5\linewidth]{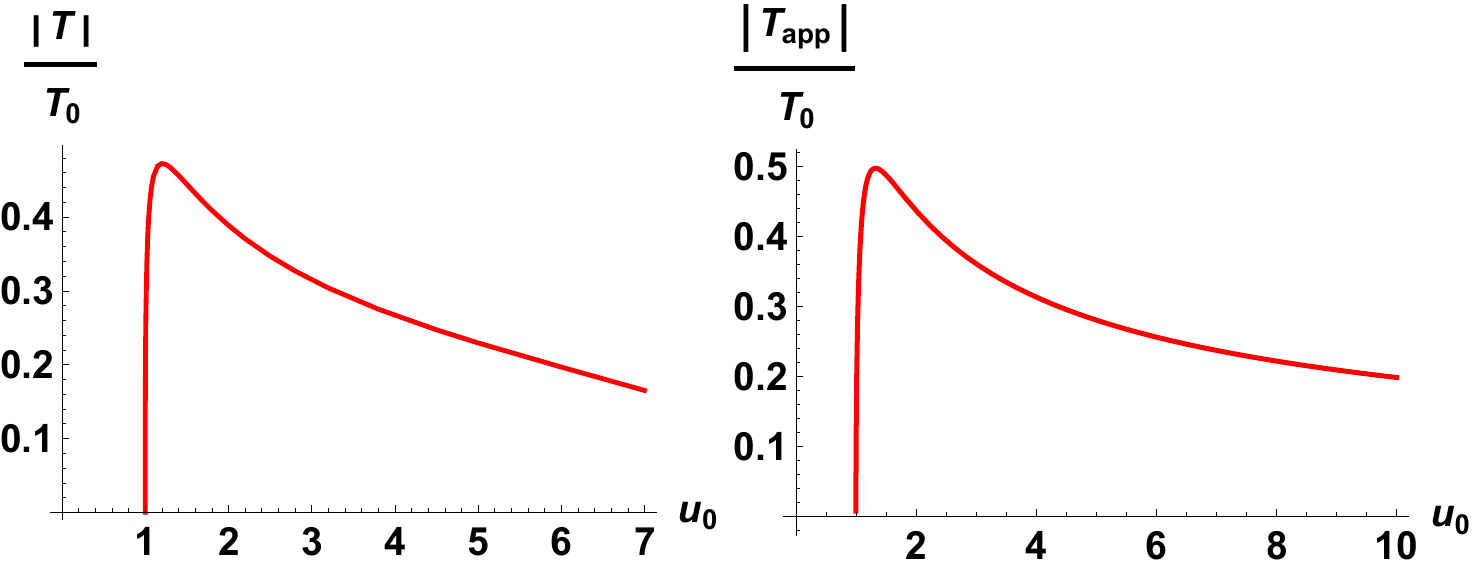}
    \caption{On the left panel, the exact time separation $|T|$ in eq.(\ref{TEED4}) in terms of the turning point $u_0$. On the right, the approximate $|T_{app}|$  in eq.(\ref{TEEappD4}) in terms of $u_0$. }
    \label{figureTu0}
\end{figure}

\begin{figure}
    \centering
    \includegraphics[width=0.8\linewidth]{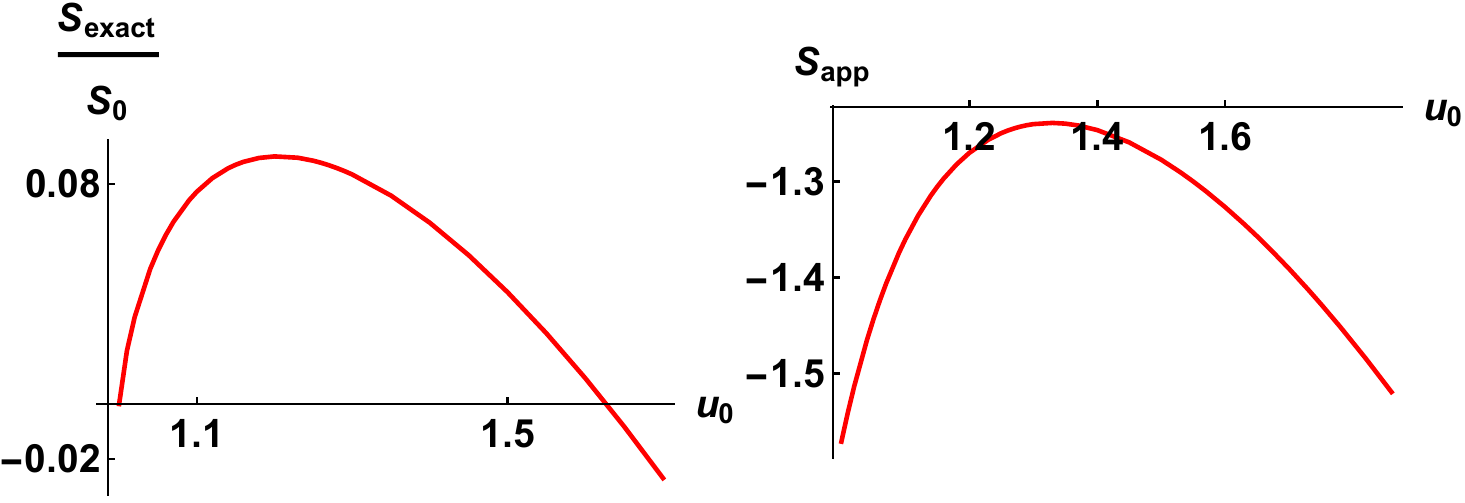}
    \caption{The exact time like entanglement entropy in terms of $u_0$, on the left. The right panel displays the approximate time-like entanglement in terms of $u_0$. Note that an integration constant can take account of the shift of both plots.}
    \label{figureSEEuo}
\end{figure}

\begin{figure}
    \centering
    \includegraphics[width=0.8\linewidth]{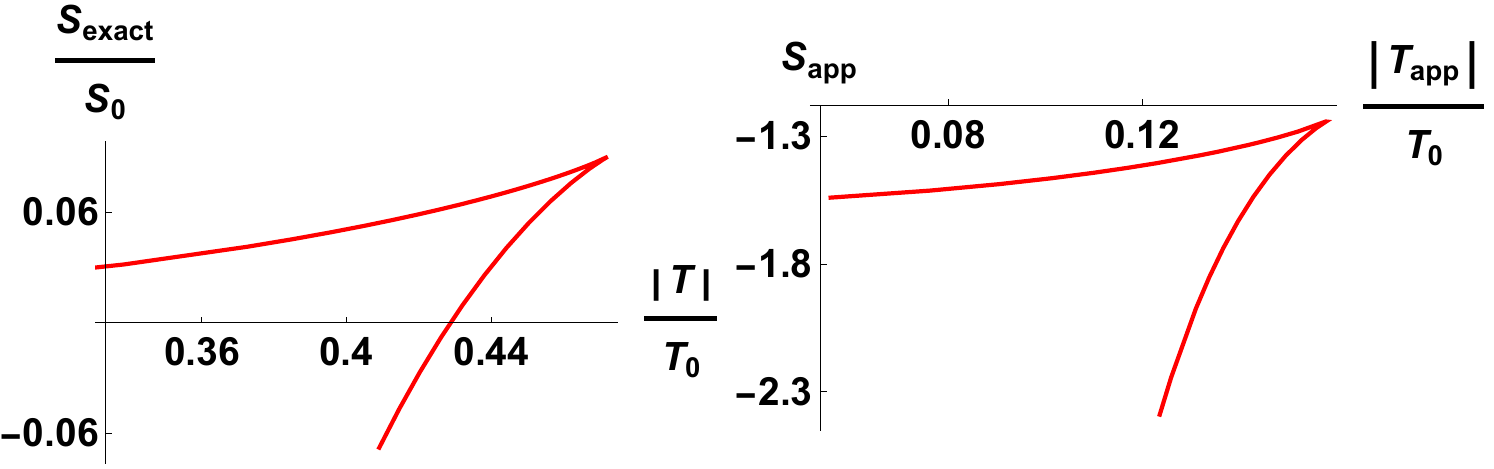}
    \caption{The exact time-like entanglement entropy in terms $|T|$ on the left, both conveniently normalised. On the right, the approximate entanglement in terms of the approximate separation (both conveniently normalised). These display the signs of a phase transition.}
    \label{sapproxtapp}
\end{figure}

We  plot the exact expressions to compare  with their analog approximate ones. Figure \ref{figureTu0} displays the exact separation $T$ in eq.(\ref{TEED4}) in terms of the turning point $u_0$ and the approximate expression in eq.(\ref{TEEappD4}). In Figure  \ref{figureSEEuo} we show the exact timelike entanglement entropy in eq.(\ref{teeD4b}) and its approximate expression in eq.(\ref{approxtEED4S1}) in terms of the turning point $u_0$. Figure \ref{figureSEEuo} displays the exact time-like entanglement entropy in terms of $u_0$ (on the left) and the parametric plot of the entropy in terms of the separation. 
Figure \ref{sapproxtapp} displays the parametric plot of $S_{app}$ in eq.(\ref{approxtEED4S1}) in terms of the approximate time separation $T_{app}$ in eq.(\ref{TEEappD4}), and the same for the exact quantities in eqs.(\ref{TEED4})-(\ref{teeD4b}).

Notice that the double-valuedness of the function $T(u_0)$ in eq.(\ref{TEED4}), or in the analog approximate quantity in eq.(\ref{TEEappD4}) indicates the presence of a phase transition. In our case, we believe this is a first order transition, as indicated by the swallow tail in Figure \ref{figureSEEuo}. 

The presence of a phase transition on the slab Ryu-Takayanagi entanglement entropy can be put in correspondence with the confining behaviour of the dual QFT. In fact, this was proposed in \cite{Klebanov:2007ws} and critically analysed in \cite{Kol:2014nqa, Jokela:2020wgs}. Based on this we propose that a phase transition in the time-like entanglement entropy  corresponds to a confining behaviour in the dual QFT. The provisos mentioned in \cite{Kol:2014nqa, Jokela:2020wgs} do apply also for the time-like entanglement entropy of \cite{Doi:2022iyj,Doi:2023zaf}. Similar ideas were put forward in \cite{Afrasiar:2024lsi}. Let us now study another confining system, with certain similarities but important differences from the one considered above.
{\subsubsection{\underline{ Anabal\'on-Ross' model for $(2+1)$ gapped theory}}}
In this case the background consist of a metric, a constant dilaton (that we take to vanish) and a Ramond five form that we do not quote here. The metric reads,
\begin{eqnarray}
& & ds^2= \frac{u^2}{l^2}\left[\lambda dt^2+dx_1^2+dv^2 + f(u) d\phi^2 \right]+\frac{l^2 du^2}{f(u) u^2}+l^2 d\tilde{\Omega}_5^2,\nonumber\\
& & d\tilde{\Omega}_5^2= d\theta^2+ \sin^2\theta d\psi^2 + \sin^2\theta \sin^2\psi\left(d\varphi_1-A_1 \right)^2+\nonumber\\
& &~~~~~~~\sin^2\theta \cos^2\psi\left(d\varphi_2-A_1 \right)^2 +\cos^2\theta \left(d\varphi_3-A_1 \right)^2.\nonumber\\
& & A_1= Q\left(1-\frac{l^2Q^2}{u^2} \right)d\phi,~f(u)= 1-\left(\frac{Ql}{u}\right)^6.\label{ARmetric}
\end{eqnarray}
For $Q=0$ the metric is AdS$_5\times S^5$. The parameter $Q$ corresponds to a VEV deformation of $N=4$
Super-Yang-Mills. The radial coordinate ranges in $Q L\leq u<\infty$. The space ends in a smooth way
if the period of the  $\phi$-direction is chosen appropriately $L_\phi=\frac{1}{3Q}$ \cite{Kumar:2024pcz}.

There is a qualitative difference between this model and Witten's in eq.(\ref{wittenmodel}). The background in eq.(\ref{ARmetric}) asymptotes to AdS$_5\times S^5$, hence the far UV is described in terms of a four dimensional CFT which is deformed and flows to a $(2+1)$ SUSY gapped and confining QFT. On the other hand,  Witten's metric asymptotes to that of $N_c$ D4 branes. Hence, the dual field theory is $(4+1)$ dimensional at high energies, which needs to be UV-completed in terms of the six-dimensional (0,2) SCFT. These differences impact on some of the physical observables as we mention below.

\begin{figure}
    \centering
    \includegraphics[width=0.8\linewidth]{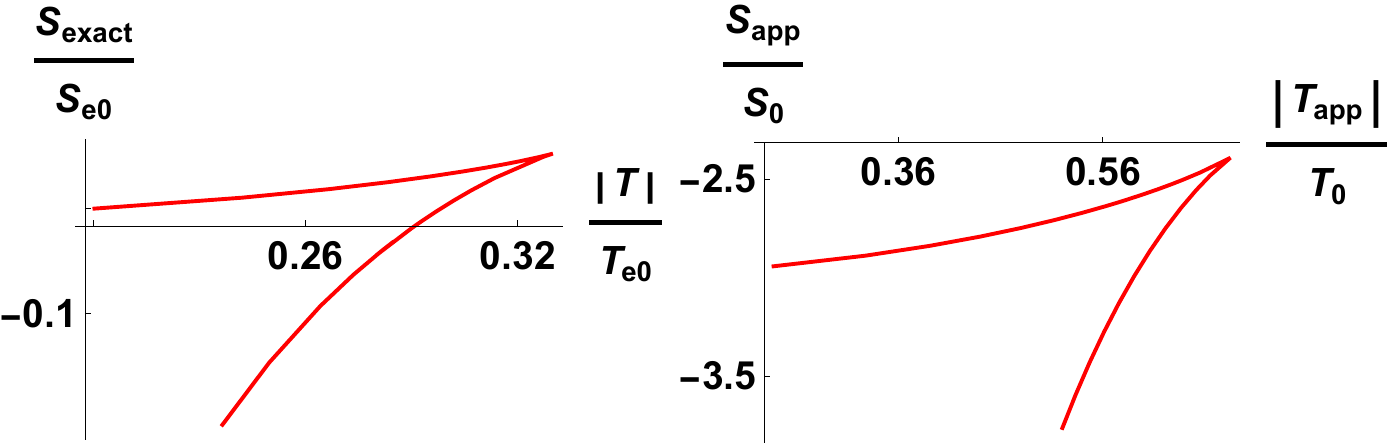}
    \caption{The exact time-like entanglement entropy for the Anabal\'on-Ross model in terms $|T|$ on the left, both conveniently normalised. On the right, the approximate entanglement in terms of the approximate separation (both conveniently normalised). These display the signs of a phase transition.}
    \label{new_fig4}
    \end{figure}

The eight manifold needed to calculate the time-like entanglement entropy is $\Sigma_8=[x_1,\phi,u,\theta,\psi,\varphi_1,\varphi_2,\varphi_3]$, with $v=0$ and $t=t(u)$. The  time-like entanglement is,
\begin{eqnarray}
 & &S^{(\lambda)}_{EE}[\hat{\Sigma}_8^{(\lambda)}]=\frac{\widehat{\cal N }}{4G_{10}}  \int_{u_0}^\infty du~ \sqrt{G^2(u) +F^2(u)t'^2},\label{tEEAR}\\
 & & G^2(u)=\frac{u^2}{l^2},~~ F^2(u)= \frac{u^6}{l^6}f(u) \lambda,~~\widehat{\cal N}=L_{x_1} L_\phi l^5\int d\tilde{\Omega}_5.\nonumber
\end{eqnarray}
We start by calculating the approximate time separation $T_{app}$, the function $Z(u)$ indicating stability of the embedding and the approximate time-like entanglement entropy $S_{app}$. In terms of a variable 
\begin{equation}
  z=\frac{u}{Q l},~~~1\leq z<\infty.  
\end{equation}
The approximate quantities read,
\begin{eqnarray}
& &T_{app}= \frac{\pi l \sqrt{\lambda}}{3 Q z_0^4}\sqrt{z_0^6-1}, ~~Z= \frac{\pi \sqrt{\lambda}}{3 Q^2 z_0^5}\frac{(4-z_0^6)}{\sqrt{z_0^6-1}},\nonumber\\
& &S_{app}= -\frac{\pi l Q^2}{6 z_0^4}(2+ z_0^6).\label{approximatesAR}
\end{eqnarray}
Here $z_0=\frac{u_0}{Ql}$ indicates how much the embedding explores the radial coordinate (large $z_0$ are small embeddings barely entering the bulk). Qualitative aspects are revealed by these approximate quantities. Notice that $T_{app}$ is double valued, indicating the possibility of a phase transition. Also, note that for $z_0$ large $T_{app}\sim z_0^{-1}$, which is the behaviour observed for CFTs in eq.(\ref{e79}). For the Witten model we obtain that at large $u_0$,
$T_{app}\sim u_0^{-1/2}$. The quantity $Z(z_0)$ is negative (indicating stability of the embedding) for $z_0>2^{\frac{1}{3}}$, hence embeddings that penetrate into the bulk deeper than this value are unstable (indicating a phase transition). The exact expressions for the time-like separation and the time-like entanglement entropy are,
\begin{eqnarray}
& &\frac{QT(z_0)}{2l}= \sqrt{\lambda} 
\sqrt{z_0^6-1}\int_{z_0}^\infty\!\! dz \frac{z}{\sqrt{(z^6-1)(z^6-z_0^6)}},\nonumber\\ 
& & \frac{S_{EE}(z_0)}{2 Q^2~ l~ \widehat{\cal N}}= \int_{z_0}^\infty
dz ~z \sqrt{\frac{z^6-1}{z^6-z_0^6}} - \int_1^\infty z dz.\label{SexactAR}\end{eqnarray}
Both integrals can be evaluated exactly in terms of Appel functions. We do not quote the result here. 
The numerical plots of the separation $T(z_0)$ and the comparison with the plot of $T_{app}(z_0)$ in eq.(\ref{approximatesAR}) are qualitatively very similar to those Figure \ref{figureTu0}. The same occurs for the  plots of  $S_{EE}$ in eq.(\ref{SexactAR}) compared with $S_{app}$ in eq.(\ref{approximatesAR}). Finally we parametrically plot $S_{app}$ in terms of $T_{app}$  and compare it with the parametric plot of $S_{EE}$ in terms of $T$. Both parametric plots are in Figure \ref{new_fig4}. Notice the similarities with the result for Witten's model in Figure \ref{sapproxtapp}. 

\section{Summary and conclusions}
Let us begin with a summary of the contents of this letter. We propose a top-down approach to computing time-like entanglement entropy, which does {\it not} rely on analytic continuation. We introduce a parameter $\lambda = \pm 1$ in front of the $g_{tt}$ component of the metric, which governs both the Euclidean and Lorentzian cases. At the conclusion of our calculations, setting $\lambda = +1$ yields the usual Ryu–Takayanagi entanglement entropy, while $\lambda = -1$ gives the result for time-like entanglement entropy. We test our method for CFTs in various dimensions and reproduce the results of \cite{Doi:2023zaf} in all cases. As a by-product, our analysis establishes a relation between the central charge of the dual CFTs (in all dimensions) and the entanglement entropy. In analogy with the Euclidean case, we propose formulas for the Liu–Mezei central charge for CFTs in various dimensions, derived from time-like entanglement entropy. We also perform this analysis for slab geometries. To illustrate our results, we discuss an infinite family of 4d $\mathcal{N}=2$ linear quiver SCFTs and their holographic duals, using them to test our expressions.

We also discuss non-conformal field theories with slab entangling regions. In these cases, the computation of time-like entanglement entropy and time separation generally requires a numerical analysis of the extremal surface. We propose analytic expressions that closely approximate the exact results. Additionally, we present a criterion for the stability of the extremal surface. These expressions are particularly useful for analyzing phase transitions in time-like entanglement entropy, especially for non-conformal QFTs and their holographic duals. In fact, using these results, we study a holographic dual to $(3+1)$-dimensional Yang–Mills theory and a supersymmetric version of $(2+1)$-dimensional Yang–Mills theory. Our approach enables the study of infinite family generalizations, which we elaborate on in \cite{NRtoappear}. We suggest that, as in the case of the Euclidean eight-manifold (corresponding to usual RT entanglement), there exists a phase transition in time-like entanglement. Specifically, the extremal surface becomes unstable beyond a certain time separation, leading to a disconnected, lower-energy configuration. In contrast, we show that such a transition does not occur for CFTs. Our approach is not addressing the important issue regarding what is the correct surface (time or space-like) that minimizes the calculation \cite{Li:2022tsv,Heller:2024whi}.

Our forthcoming paper \cite{NRtoappear} provides detailed derivations of several results presented in this letter and extends them to infinite families of CFTs and confining models in various dimensions. See, for instance, the models discussed in \cite{Akhond:2021ffz, Lozano:2019zvg, Lozano:2020bxo, Legramandi:2021uds, Fatemiabhari:2024aua, Maldacena:2000yy, Conde:2011aa, Nunez:2023xgl, Nunez:2023nnl}.

We plan to continue developing this promising line of investigation.

\bibliographystyle{unsrt}
\bibliography{refs}

\noindent \textit{Acknowledgments:} We thank Dimitrios Giataganas and Tadashi Takayanagi for useful and interesting comments. DR would like to acknowledge The Royal Society, UK for financial assistance. DR also acknowledges the Mathematical Research Impact Centric Support (MATRICS) grant (MTR/2023/000005) received from ANRF, India. C. N. is supported by STFC’s grants ST/Y509644-1, ST/X000648/1 and ST/T000813/1.

\end{document}